# Inference of gene loss rates after whole genome duplications at early vertebrates through ancient genome reconstructions

Haiming Tang[1,*],  Angela Wilkins[1]
1 Mercury Data Science, Houston, TX
* Corresponding author

## Abstract

The famous 2R hypothesis was first proposed by Susumu Ohno in 1970. It states that the two whole genome duplications had shaped the genome of early vertebrates. The most convincing evidence for 2R hypothesis comes from the 4:1 ratio chromosomal regions that have preserved both gene content and order in vertebrates compared with closely related. However, due to the shortage of such strict evidence, the 2R hypothesis is still under debates.

Here, we present a combined perspective of phylogenetic and genomic homology to revisit the hypothesis of 2R whole genome duplications.  Ancestral vertebrate genomes as well as ancient duplication events were created from 17 extant vertebrate species. Extant descendants from the duplication events at early vertebrates were extracted and reorganized to partial genomes. We then examined the gene order based synteny, and projected back to phylogenetic gene trees for examination of synteny evidence of the reconstructed early vertebrate genes. We identified 7877 ancestral genes that were created from 3026 duplication events at early vertebrates, and more than 50% of the duplication events show synteny evidence. Thus, our reconstructions provide very strong evidence for the 2R hypothesis.

We also reconstructed the genome of early vertebrates, and built a model of the gene gains and losses in early vertebrates. We estimated that there were about 12,000 genes in early vertebrates before 2R, and the probability of a random gene get lost after the first round of whole genome duplication is around 0.45, and the probability of a random gene get lost after the second round of whole genome duplication is around 0.55.

This research provides convincing evidence for the 2R hypothesis, and may provide further insights in vertebral evolution.

**Data availability:** https://github.com/haimingt/Ohnologs-and-2R-WGD

## Introduction

The hypothesis that the vertebrate genome underwent two whole genome duplications (2R WGDs) in early vertebrates was first proposed by Susumu Ohno in 1970(Ohno 1970a). The best-known evidence for 2R hypothesis is the case of 4 HOX clusters on different chromosomes in HUMAN and other vertebrates and only 1 HOX cluster in invertebrates that are closely related with vertebrates like cephalochordate amphioxus (Garcia-Fernandez 2005; Lemons and McGinnis 2006; Lundin 1993). In late 1990s and early 2000s, the hypothesis was hotly debated (Makalowski 2001). In recent years, with the development of genomic sequencing and accumulated evidences, most researchers have accepted the 2R hypothesis (Cañestro 2012; Kasahara 2007; Panopoulou and Poustka 2005). Yves Van de Peer stated that "2R or not 2R is not the question anymore" in 2010 (Van de Peer, et al. 2010). Only few scientists still hold opponent opinions (Abbasi 2015, 2010). Opponents reject 2 rounds of whole genome duplications while attribute the expansion of vertebral genomes to a series of small-scale segmental duplications (Abbasi 2015, 2010; Friedman and Hughes 2001, 2003).

In theory, if two rounds of genome duplication (2R-WGD) had occurred, we would expect the presence of closely linked sets of paralogous genes on 4 chromosomal regions in vertebrates whereas closely-related invertebrates contain only one set of linked genes (Van de Peer 2004). In addition, these paralogous genes should be dated to the time periods of early vertebrates. Also the paralogous genes should cover a large portion of the genome of early vertebrates. The well-known HOX gene clusters satisfy the criteria. There are four copies of HOX gene clusters which consist copies of HOX genes as well as other adjacent genes in 4 different chromosomes of *Homo Sapiens* and other land vertebrates compared with only 1 copy in fish-like chordate *Cephalochordate Amphioxus*. Other similar gene clusters reported include MHC (Kasahara 2007; Katsanis, et al. 1996) and the EGF ligands (Laisney, et al. 2010). These paralogons provide convincing evidences for 2R hypothesis. However, these examples only cover a very small portion of the genome. A careful examination of additional supporting evidences for 2R hypothesis shows that they may not be sufficient to ultimately prove the 2R hypothesis.

Hughs *et al* revealed that less than 5% of homologous gene families follow the 4:1 rule through comparison of human and drosophila genomes (Friedman and Hughes 2003). Besides, only limited conservation of local gene order and no 4:1 ratio paralogons between vertebrates and Florida lancet were found (Putnam, et al. 2008). The Synteny Database (http://teleost.cs.uoregon.edu/synteny_db) predicts 231 paralogy clusters with more than 5 genes in human using Florida lancet as the out-group, these clusters cover less than 15% of the Human genome (Cañestro 2012; Catchen, et al. 2009).

Researchers thus relaxed the strict gene-order criteria and use content-based synteny (Abi-Rached, et al. 2002; Hampson, et al. 2005; Vandepoele, et al. 2004) as evidence of 2R hypothesis. Putnam et al found extensive conservation of gene linkage on the scale of whole chromosomes, which they named as "macro-synteny"

(Putnam, et al. 2008). Through human, fly and nematode genomes, Aoife McLysaght et al. found 504 paralogons with 3 or more distinct duplicated genes in each paralogon which covered 79% of the genome (McLysaght, et al. 2002) . Dehal and Boore's research has been widely accepted as a convincing evidence for 2R hypothesis (Dehal and Boore 2005). They extracted paralogs that originated from duplications predating the fish-tetrapod split and mapped their positions in the human genome. Using a sliding window that is 50 genes to the left and 50 genes to the right of a query gene, "hits" were obtained if corresponding windows in other chromosomes contain the query gene's paralogs. They found that 4-fold matching windows covered the clear majority of the Human genome. However, 4-fold matching windows are much less convincing as 4:1 ratio paralogons. It doesn't show evidence of linkage between distinct genes within a same sliding window. Besides, a sliding window is compared with windows in other chromosomes with similar locations; thus, it neglects the paralogous genes outside the sliding windows. Recently, Param Singh et al investigated the conservation of content-based gene synteny of six amniotic vertebrates relative to six invertebrate out-group genomes using an algorithm that integrates the synteny information from both self and out-group comparisons (Singh, et al. 2015). Under their relaxed criteria, they identified 7831 human Ohnologs from 2642 Ohnolog families, out of which 96.7% have 4 or fewer genes. These genes are spread out in different chromosomes and consist a large portion of the Human genome. Thus, it provides a strong evidence for 2R hypothesis.

In summary, although there are lots of evidences supporting the 2R hypothesis, conclusive evidences that 4:1 ratio paralogons with conserved both gene content and gene order cover the majority part of an ancestral genome of early vertebrates are still lacking. As pointed out by Masanori Kasahara (Kasahara 2007), "investigators who identified paralogons by map-based approaches were the proponents of the 2R hypothesis, whereas phylo-geneticists who analyzed paralogs by tree-based approaches were the opponents." The opponents of the 2R believe the duplications in early vertebrates are attributed to a series of regional duplications. Hugh et al and Abbasi et al argued that paralogs A-D generated by two rounds of WGD should display the tree topology (AB) (CD) (Abbasi 2015; Friedman and Hughes 2001, 2003), while a substantial majority of genes in paralogons do not have this pattern.

Here, we developed an unprecedented perspective to revisit the 2R hypothesis. The method is a well-balanced combination of phylogeny and strict gene-order based genome homology: Firstly, we extracted genes that are originated from duplications just predating the fish-tetrapod split of 17 current vertebrates and related Florida lancet gene from PANTHER database (Mi, et al. 2017; Mi, et al. 2016). Extraction of these genes that are duplicated in early vertebrate could greatly help eliminate the distraction of genes that are duplicated at other periods. Secondly, we reshaped these genomes via keeping only the genes extracted above and then detected within and between genomic paralogons that preserved gene orders using i-ADHoRe 3.0 (Proost, et al. 2012). Finally, we summarize the evidence of synteny and determine if synteny evidence exists between any 2 branches of a duplication node at early vertebrates. Thus, our analysis yields the percentage of duplications events which is from "segmental" duplications.

We also reconstructed the gene repertoire of early vertebrates: including the early vertebrates genes that were inherited from its ancestor and were not duplicated as well as the ancestral genes in early vertebrates that duplicated with 2, 3, 4 or more copies. Detailed methods are in our previous paper (Tang, et al. 2018). Using these reconstructions, we were able to construct a model of gene losses during and right after the hypothesized 2 rounds of whole genome duplications. In brief, we set the total number of genes in early vertebrates before 2R is $N$; In the first round of whole genome duplication, the genomes expands to $2*N$ genes; Then a random gene gets lost after the first round of whole genome duplication with probability $p1$; Following is the second round of whole genome duplication which duplicates all existing genes after the first round of loss; Finally a random gene get lost after the second round of whole genome duplication with probability $p2$. Specifically, if there is only 1 copy of paralogs left in the genome, the loss probability is set to $p0$. We explore the estimates of $N$, $p0$, $p1$ and $p2$ that best fit our reconstructions using a simulation study.

## Methods

### Ancestral reconstructions

We performed extensive ancestral reconstructions from the universal last common ancestor in our previous paper. Detailed protocols could be found there. The ancestral reconstructions were performed on PANTHER database. PANTHER is a large collection of protein families that have been subdivided into functionally related subfamilies (Mi, et al. 2017; Mi, et al. 2016). Hidden Markov models (HMMs) are built for each family and subfamily for classifying additional protein sequences. The PANTHER Classifications are the results of human curation as well as sophisticated bioinformatics algorithms. For each PANTHER family, a PANTEHR tree was created with reconciliation with a predefined species tree. In this analysis, the PANTHER database version 10 (release date July 2016) contains more than 1 million genes from 104 genomes, including 17 vertebrates and closely related invertebrates like Florida lancelet. The phylogenetic tree is included in Supplemental material Part 1.

Ancestral reconstructions of the internal nodes of each phylogenetic gene tree in PANTHER are classified to 2 types: speciation and duplication. For a speciation node, it is enforced to form a monophyletic clade, and a ceancestors is assigned to the node by finding the common ancestor of the descendant species through tracing a reference species tree. Thus, the speciation node represents an ancestral gene in the ceancestor that had existed in evolutionary history. A duplication node is inferred only when a given speciation node contains more than one gene from the same species (within-species paralogs). For ancestral duplication node, it is inferred when more than one ancestral gene of the same ceancestor (or a younger ceancestor, as the older ceancestor must have existed in evolutionary history for the young ceancestor to inherit the gene from) is found. Thus, the age of a duplication node is inferred by the oldest ceancestor in its direct descendants.

**Extraction of genes originated from duplications just predating the fish-tetrapod split**

To extract the genes that are originated from duplications at early vertebrates, each PANTHER phylogenetic tree is searched for duplication nodes that are dated to early vertebrates periods. At least one branch should contain genes of both fish and tetrapod, and no branch should contain genes of non-vertebrates species. Illustrated in Figure 1.I. These duplication nodes are identified as duplications events that have happened at early vertebrates. We then extract all descendant genes from such early vertebrates duplication nodes. By this restriction step, we are able to exclude genes that are duplicated at time periods earlier than early vertebrates as well as genes that do not have duplicated homologs. However, genes that are duplicated at time periods later than early vertebrates, like fish-specific duplications, primate specific duplication … will also be included. The problem lies in an unsolved scientific problem that in a duplication event, one or more duplicated copies are gained from copying the original gene. After duplication, these descendants are "identical" except their genomic locations, thus the "original copy" and the "duplicated copy" could not be effectively differentiated. Some researchers are even against of the term "original copy". Their belief is that, the true original gene will disappear after the duplication events leaving 2 duplicated copies. To avoid confusion, we use "original copy" here to represent the gene copy that has stayed in the same genomic region. In some cases, the "original copy" of a later duplication event may be the "duplicated copy" of the duplication events at early vertebrates. To avoid delimiting these "original copies" duplicated from the proposed 2R WGDs, we extract all descendant genes from duplication nodes at early vertebrates for later analysis.

**Synteny detection using i-ADHoRE 3.0**

Extracted genes that originated from early vertebrates duplications are mapped to their chromosomal locations using Ensembl Biomart version 84 (Yates and Akanni 2016). For each genome, genes are sorted by their locations in individual chromosomes, and lists of these extracted genes are created for each chromosome. For unassembled genomes, lists of genes are created for each scaffold. The ranked lists of extracted genes from the 17 vertebrates and 1 closely related invertebrate genomes are in Supplemental Material Part1. Paralogous relationships of genes are labeled not only by PANTHER families they belong to, but also by the duplication node at early vertebrates. That's because large PANTHER families often have many duplication nodes at the same ceancestor. For example, in the PANTHER family for HOX gene clusters, different duplication nodes represent different HOX genes. Thus treating the duplication nodes in the same PANTHER family differently could ensure more accuracy.

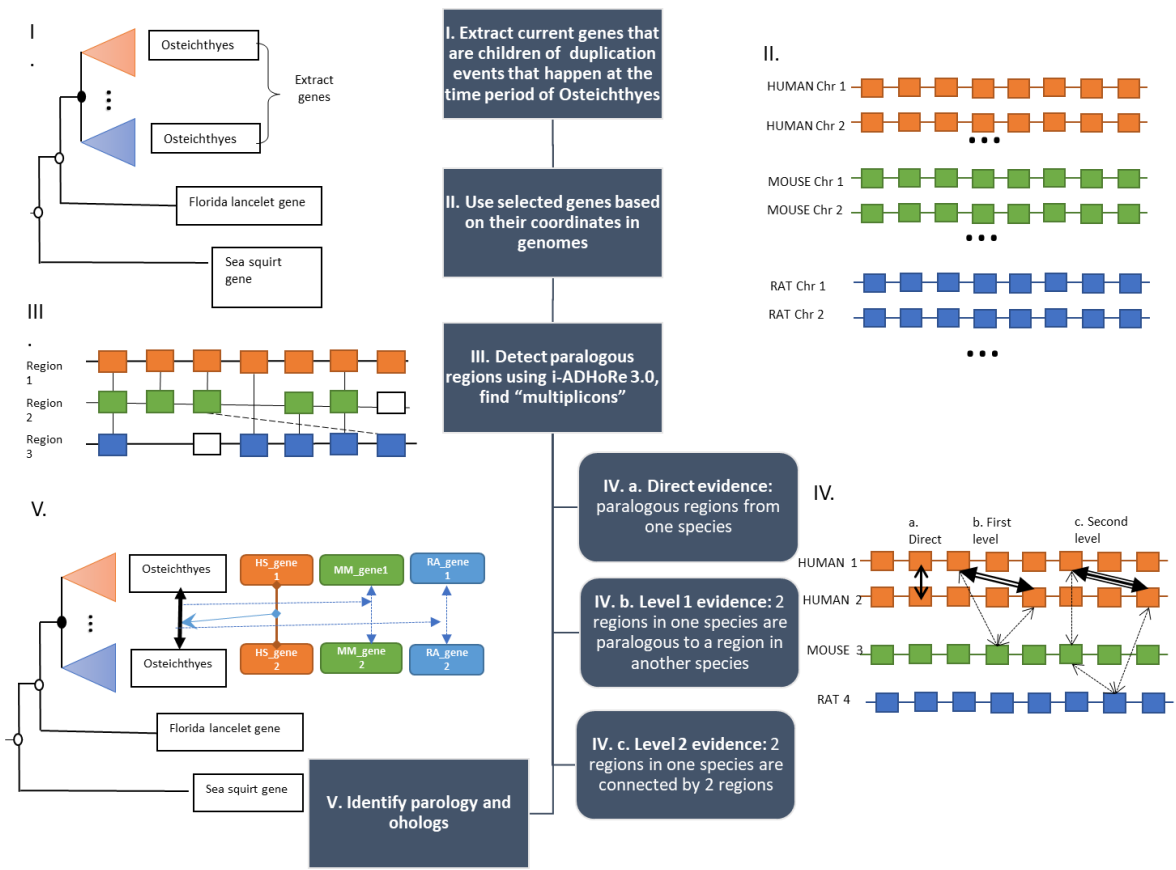

**Figure.1 Flowchart for detection of synteny evidence for branches of duplication nodes at early vertebrates**

*I. Extract extant genes that are children of duplication events that happen at the period of Osteichthyes in PANTHER 10. Sub-figure I on the upper left shows part of a phylogentic tree of PANTHER 10. The orange and blue triangles that are labeled as "Osteichthyes" represent collapsed branches of extant vertebrate genes. The black dot that connects the "Osteichthyes" branches is a duplication node. It is dated to a time-period after divergence from invertebrate species florida lancelet, and before divergence to fish and land vertebrates. II. The extant genes in Osteichthyes branches are from 17 verterbrates: Human, Chimpanzee, Mouse, Rat, Fish, Fugu Fish, Chick, Cat, Dog, Green anole, Bovin, Zebrafish, Rhesus macaque, Gray short-tailed opossum, Duckbill platypus, Japanese pufferfish and Western clawed frog. The extracted genes form lists of genes based on their coordinates on chromosomes or scaffolds. Sub-figure II show several lists, rectangles represent individual genes. III. Paralogous regions from lists of extracted genes are detected using i-ADHoRe 3.0. Sub-figure III show a "multiplicon" detected, the vertical lines that connects genes in different regions are homologous genes. Paralogous regions that form a "multiplicon" have homologous genes that also preserve the gene orders. IV. Evidence of synteny for homologous gene pairs of the same species are summarized from "multiplicons". Sub-figure IV show 3 levels of evidences: direct evidence: 2 genes are in a multiplicon from the same species; level 1 evidence: 2 regions in one species are paralogous to a region in another species; level 2 evidence: 2 regions in one species are connected by 2 regions. V. Synteny evidence between 2 Osteichthyes branches is summarized by synteny evidence of descent genes in the branches. Sub-Figure V shows that human gene 1 from one branch has synteny evidence with human gene2 in another branch (HS_gene1 and HS_gene2 connected by vertical line), thus we infer synteny evidence for these 2 branches (connected by vertical bold line), then for homologous gene pairs from these 2 branches, if they do not have evidence of*

*synteny from Step IV, we infer them to have synteny evidence. Thus homolgous gene pairs from branches with evidence of synteny are inferred as "ohnologs".*

i-ADHoRE 3.0 (Proost, et al. 2012) is a software package for detection of genomic regions which are statistically significantly conserved with both gene content and order. Briefly, it works by first detecting initial pairs of homologous segments, and aligning them to form a profile with combined gene order and content information. The profile is then used to detect additional homologous segments, which will be added to the profile as well. And the search is repeated using the updated profile until no additional segments can be found.

We then apply i-ADHoRE 3.0 to detect synteny regions using the ranked gene lists of all 17 vertebrate's genomes and the Florida lancet genome described above. Default parameters are used. Both within and among genomic syntenies are detected. The collinear mode that considers both content and gene order was used. Minimum number of matched paralogous genes between 2 segments was chosen to be 3 to allow a more relaxed detection of synteny regions.

**Synteny evidences between 2 paralogous genes**

"Multiplicon" is the term introduced by i-ADHoRE 3.0. It represents 2 homologous segments consisting of homologous genes with conserved orders. For paralogous gene pairs in the same species, we summarize 3 levels of synteny evidences: direct evidence, level 1 evidence and level 2 evidence. As Illustrated in Figure 1.IV, direct evidence between 2 paralogous genes is summarized from the scenario that both genes are found in the same multiplicon whose genomic regions are from the same species or 2 different species. Level 1 evidence between 2 paralogous genes is from the condition that 2 genes have no direct evidence, but both genes have direct evidence with a third gene. Level 2 evidences between 2 genes is from the condition that 2 genes have no direct or level 1 evidence, and the 2 genes are connected by 2 intermediate genes: gene A has direct evidence with gene B, gene B has direct evidence with gene C, and gene C has direct evidence with the gene D, thus gene A and gene D have level 2 evidence of synteny. We extend the direct evidence to level 1 and level 2 evidences because multiplicons and paralogous gene pairs may fail to be detected due to gene losses or change of gene orders. The multiplicons together with the paralogous gene pairs with the 3 levels of synteny evidence are summarized in supplemental material part 2.

**Synteny between any 2 branches of a duplication node at early vertebrates**

In PANTHER gene trees, a duplication node defines a duplication event at a ceancestor. Each descendent branch represents a copy of an ancestral gene duplicated in the corresponding duplication event. Thus, descendant branches of a duplication node at early vertebrates represent gene copies that were gained through duplication in ancestral early vertebrates. Evidence of synteny for the inferred ancestral genes descendental from duplication at early vertebrates indicats that these descendant genes are from ancestral segmental duplications instead of tandem duplications. If most of the duplication events at early vertebrates are from segmental duplications, we could safely infer whole genome duplications. As illustrated in Figure 1.V, we infer the synteny between ancient duplicates by summarizing the synteny evidences of extant species genes in descendants. If human gene 1 (HS_gene1) in branch 1 and human gene 2 (HS_gene2) in branch 2 are paralogous genes in a multiplicon or have level 1 or 2 evidences of synteny, then we conclude that the 2 inferred ancestral genes represented by branch 1 and branch 2 have evidence of synteny. For any 2 branches, we extract paralogous gene pairs of the same vertebrate species, one from each branch. If any of the same species gene pairs from 2 different branches of the same duplication node at early vertebrates has synteny evidence, we conclude that there is evidence of synteny between these 2 branches. The rationale behind the evidence extension is the hypothesis that genes frequently get lost and chromosomes often undergo rearrangements and the order of genes may change during evolution.

**Simulation Study to estimate gene loss rates after whole genome duplications**

We have reconstructed the gene repertoire of early vertebrates: including the early vertebrates genes that were inherited from its ancestor and were not duplicated as well as the ancestral genes in early vertebrates that duplicated with 2, 3, 4 or more copies. Detailed methods are in our previous paper (Tang, et al. 2018). We were able to reconstruct the number of "ancestral genes" that get lost before divergence of early vertebrates N0; the number of "ancestral genes" that are inherited from their ancestor and are not duplicated at early vertebrates N1; the number of duplication nodes at early vertebrates that have 2 branches N2; the number of duplication nodes that have 3 branches; and the number of duplication nodes that have 4 branches N4. Using these reconstructions, we were able to construct a model of gene gains and losses during and right after the hypothesized 2 rounds of whole genome duplications. Illustrated in Figure 2. We set the total number of genes in early vertebrates before 2R is *N*; In the first round of whole genome duplication, the genomes expands to 2\**N* genes; Then a random gene gets lost after the first round of whole genome duplication with probability *p1*; Following is the second round of whole genome duplication which duplicates all existing genes after the first round of loss; Finally a random gene get lost after the second round of whole genome duplication with probability *p2*. Specifically, if there is only 1 copy of paralogs left in the genome, the loss probability is set to *p0*. We explored all possible combinations of *N*, *p0*, *p1* and *p2* to find best estimates that fit our reconstructions: *N* from 12,000 to 20,000; *p1* from 0.05 to 0.95; *p2* from 0.05 to 0.95; *p1* from 0.01 to 0.99. We then compare the estimations from the various combination of parameters with the duplication statistics at early vertebrates N0, N1, N2, N3, N4 numbers described above, and select the best parameters N, p0, p1 and p2.

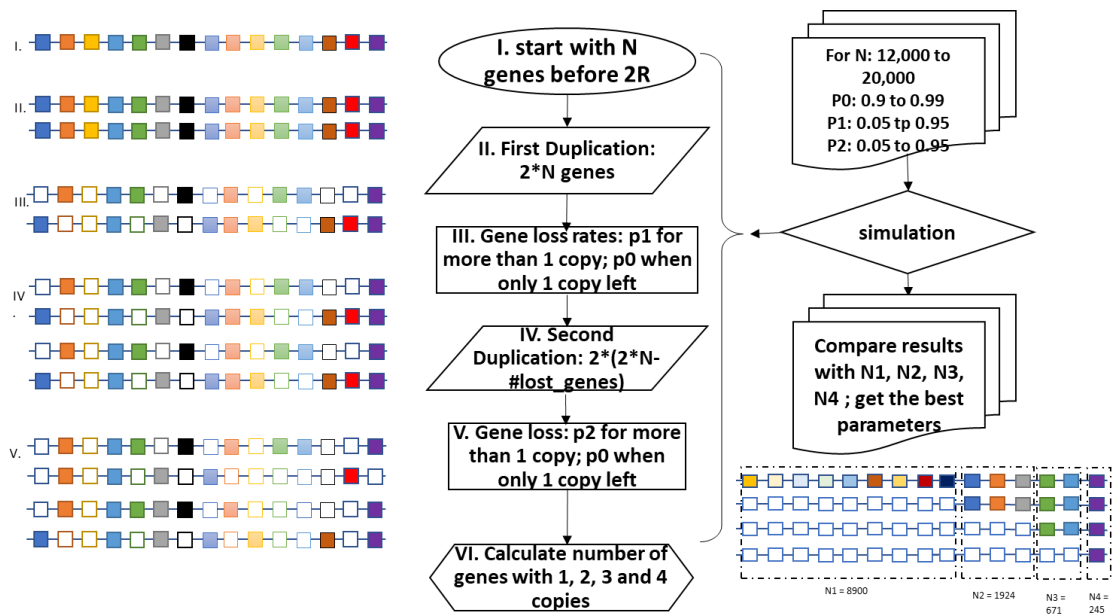

**Figure 2 Simulation study for gene loss rate after whole genome duplications**

*The chain of colored boxes at I. represent a list of N genes before the 2 rounds of whole genome duplications. During the first round of whole genome duplication II., this list of N genes gets duplicated, and we get 2 identical lists of genes. Then we do the simulation for genes in each list to randomly get lost. The blank boxes in III represent lost genes. Then comes the second round of whole genome duplication. The remaining genes in III get duplicated, and we get 4 lists of genes. We repeat the simulation process for genes to randomly get lost after the second round of duplication. And finally, we compare the simulation results with observed reconstruction results which are shown in the right bottom. Specially, we have 4 numbers: N1 = 8900, N2= 1924, N3=671, N4 =245. N1 is the reconstructed ancestral gene at early vertebrates (branch Osteichthyes in PANTHER phylogenetic trees). N2 is the number of duplication nodes at early vertebrates which have 2 branches, each branch is a speciation node which stands for one copy of ancestral gene at early vertebrates. N3 and N4 are the numbers of duplication nodes at early vertebrates with 3 branches and 4 branches separately.*

*For first round of duplication, all N genes get a duplicated copy. Then a random number between 0 and 1 is generated for each of 2*N genes, if the random number is larger than p1, then it is deleted from the gene pool. For a gene pair consisted of the original copy and the duplicated copy, if one of the copies has already been deleted, the other copy is lost only if the random number is larger than p0. For the second round of duplication, all genes that are left get an extra duplicated copy. Similar loss procedure using "p2" and "p0" is applied. In the end, we count the total number of N genes that have 0, 1, 2, 3 and 4 copies:* n0*,* n1*,* n2*,* n3 *and* n4*.*

*We run the simulations with N from 12000 to 20000, p0 from 0.01 to 0.99, p1 from 0.05 to 0.95 and p2 from 0.05 to 0.95, and generate sets of* n0*,* n1*,* n2*,* n3 *and* n4 *for different combinations of parameters. In our paper "Reconstruction of gene gain and loss since the universal last common ancestor", we estimate the number of "ancestral genes" for early vertebrates: the number of "ancestral genes" that get lost before divergence of early vertebrates N0; the number of "ancestral genes" that are inherited from their ancestor and are not duplicated at early vertebrates* N1*; the number of duplication nodes at early vertebrates that have 2 branches* N2*; the number of duplication nodes that have 3 branches; and the number of duplication nodes that have 4 branches* N4*. We compare the sets of* n0*,* n1*,* n2*,* n3 *and* n4 *to the reconstructed* N0, N1, N2, N3, N4 *numbers, and select the best parameters* N, p0, p1 *and* p2*.*

## Results

## Summary of the synteny evidences of duplication nodes at early vertebrates

The synteny evidences of all duplication nodes at early vertebrates are summarized in Table 1. We have found a total of 3026 duplication nodes at early vertebrates, 1924 of them have 2 branches, 671 of them have 3 branches, and 8.1% (245) of them have 4 branches. These branches comprise the 4851 copies of "early vertebrate" genes that were gained from duplication events. We also estimated the inherited "early vertebrate" genes from assembling evidence from PANTHER phylogenetic gene trees. Basically, each speciation node at early vertebrates ("Osteichthyes" as used in PANTHER) accounts 1 for inherited early vertebrate gene. We estimate that in early vertebrates common ancestor, there are 8900 inherited genes at the period of early vertebrates, and 3026 duplication events, 1924 of these yield 2 copies, 671 yield 3 copies and 245 with 4 copies, forming a total of 7877 gene copies from duplication. Thus, the duplication events at early vertebrates increase the total number of genes by 40.67% from 11926 (8900 inherited genes plus 3026 genes that have evidence of duplications) before the hypothesized 2R to 16777 (11926 plus the 4851 new gene copies that are duplicated from the 3026 genes after 2R.) Besides, 25.37% of the ancestral genes in early vertebrates are duplicated (3026 over 11926). Although only 8.1% of the duplication nodes have 4 branches, 79% of all them have 4 branches or less, thus the results could potentially be explained by gene losses followed by the hypothesized 2 rounds of whole genome duplications.

This finding is also consistent with previous research: most of the Ohnolog families have 2 branches, followed by 3 branches, and most families have branch size of 4 or smaller. Hughs et al found that less than 5% of homologous gene families follow the 4:1 rule through comparison of human and drosophlia genomes (Hughes 1999); Singh et al discovered 2642 Ohnolog Families, only 9.3% (245) of them have size 4 (Singh, et al. 2015).

Overall, 57.5% (1740/3026) of the duplication nodes at early vertebrates show some evidence of synteny. The percentage of duplication nodes with evidence of synteny is perhaps over estimated because duplication nodes with more branches have higher total number of comparable branch pairs. For example, a duplication node of 3 branches: branch A, B and C has 3 comparable pairs A-B, A-C and B-C, while a duplication node of 4 branches have 6 possible pairs. After adjusting for this, only 40% of all possible branch pairs from the duplication nodes at early vertebrates have evidence of synteny. The total number of branches with synteny evidences with other branches is 4441, 56.3% of a total number of 7877 branches. Bedsides duplication nodes with more branches have higher probability of synteny evidence: for duplication nodes of 4 branches, 78.3% have synteny evidence.

The interpretation of the synteny evidence is that around 57% of all the duplication events that have happened at the specific time periods of early vertebrates are from duplications of chromosome segments that contain at least 3 consecutive genes. Combined with the results that 25.37% of the ancestral genes in early vertebrates are duplicated, and that these duplication events increase the total number of genes by 40.67%, our study provides a very strong evidence for the 2R hypothesis.

**Table 1 Summary of the synteny evidences of duplication nodes at early vertebrates**

| # of branches | Total # Duplication nodes | Has synteny evidence | Some evidence | All evidence | Total # direct | Total # level 1 | Total # level l2 | Total # pairs | # evidences | Total # branches with synteny evidence |
|---|---|---|---|---|---|---|---|---|---|---|
| 2 | 1924 | 899 (46.7%) | 0 | 899 | 476 | 360 | 63 | 1924 | 899 (46.7%) | 1798 |
| 3 | 671 | 474 (70.64%) | 312 | 162 | 402 | 368 | 103 | 2013 | 873 (43.4%) | 1185 |
| 4 | 245 | 192 (78.3%) | 166 | 26 | 214 | 262 | 93 | 1470 | 569 (38.7%) | 597 |
| 5 | 80 | 73 (91.3%) | 71 | 2 | 90 | 120 | 48 | 800 | 258 (32.3%) | 259 |
| > 5 | 106 | 102 (96.2%) | 101 | 1 | 198 | 533 | 220 | - | - | 602 |

**Estimate the gene loss rate after the whole genome duplications**

By assuming the correctness of the 2R hypothesis, we performed a simulation study to estimate the total number of genes in early vertebrates: *N*, the probability of a random gene gets lost after the first round of whole genome duplication: *p1*, the probability of a random gene get lost after the second round of whole genome duplication: *p2*, and the probability of a random gene get lost if there is only one copy of this gene left in genome: *p0*. We did a series of simulations that explore all combinations of parameters: *N* from 12,000 to 20,000; *p1* from 0.05 to 0.95; *p2* from 0.05 to 0.95; *p1* from 0.01 to 0.99. The limit 12,000 is chosen because it is the minimum total number of ancestral genes in early vertebrates through our reconstruction. Each simulation generates 4 lists of homologous genes with some genes get lost, we then count the homologous gene pairs with 0, 1, 2, 3, 4 copies of genes, and compare with the results we reconstruct and assemble from PANTHER phylogenetic gene trees. Through the reconstruction, we estimate there are 8900 genes without duplicated copy, 1924 duplication nodes with 2 copies, 671 duplication nodes with 3 copies, and 245 duplication nodes with 4 copies, and 1000 genes that get lost in ancestral early vertebrates after the hypothesized 2R whole genome duplications and before the divergence of fish and land vertebrates.

The simulation study yields the parameters that fit observed reconstructions. We list all potential parameter combinations in Table 2. In summary, we estimate there are a total of about 12000 genes in early vertebrates, the probability that a random gene gets lost after the first round of whole genome duplication is around 0.45, the probability that a random gene gets lost after the second round of whole genome duplication is around 0.55, and the probability that a random gene gets lost if there is only one copy of this gene left in genome is 0.09. Unlike the previous theory that "vast majority" of the duplicated genes are lost immediately after the whole genome duplications(Albalat and Canestro 2016), our simulation yields much smaller loss rates. This finding could help address the objections to the 2R that instead of massive gains by whole genome duplications followed by massive losses immediately after that, only small-scale duplications had happened, without such massive losses.

Duplications add extra copies of genes to existing genome, and if the duplicated gene copies do not have a role in genome functioning, they could get lost very quickly in evolution process due to negative selection. We estimate that 45%~55% percentage of duplicated copies of genes get lost. The remaining duplicated copies are likely to have played important roles in shaping early vertebrates.

The gene loss model is a first attempt for modeling the gene losses and during and right after the 2R whole genome duplications. Several other factors could affect the simulation: tandem duplications could add extra copies of genes next to existing genes locations in the genome: this may help explain the duplication nodes at early vertebrates with 5 or more descendant branches. Genes could get lost during the time periods after the second round of whole genome duplication till the extant species: that may address why the estimated loss rate after the second round of whole genome duplications is slightly larger than that after the first round.

Discussions

Comparison with Ohnologs reconstructed by Singh et al.

Singh et al. has discovered a total number of 7831 human Ohnologs forming 2642 families using their relaxed criteria (Singh, et al. 2015). Out of these families, we found 148 families whose genes fall into different PANTHER family and 588 families whose genes are not originated from duplications at early vertebrates. After eliminating these families, we found 146 families whose genes are descendants of 2 or more duplication nodes in one PANTHER family. Thus, after excluding the families above, 1760 (66%) families have consistent phylogenetic tree structure with PANTHER 10. These 1760 families are directly comparable with duplication nodes at vertebrates. There are 1175 families with 2 groups, 438 families with 3 groups, 135 families with 4 groups, and 12 families with more than 4 groups. Each Ohnolog family is matched to a corresponding duplication node in PANTHER and compared with 3 conditions: comparable results in both research, Singh et al's

**Table 2 Simulation results of gene loss rate after whole genome duplications**

| N | p0 | p1 | p2 | N0 | N1 | N2 | N3 | N4 |
|---|---|---|---|---|---|---|---|---|
| 12000 | 0.1 | 0.3 | 0.6 | 808 | 7266 | 2896 | 882 | 148 |
| 12000 | 0.09 | 0.45 | 0.55 | 924 | 7264 | 2920 | 736 | 156 |
| 12000 | 0.08 | 0.45 | 0.55 | 833 | 7406 | 2867 | 742 | 152 |
| 12000 | 0.07 | 0.45 | 0.55 | 740 | 7426 | 2961 | 729 | 144 |
| 12000 | 0.07 | 0.3 | 0.6 | 568 | 7388 | 2949 | 928 | 167 |
| 12100 | 0.09 | 0.3 | 0.6 | 817 | 7360 | 2943 | 827 | 153 |
| 12100 | 0.06 | 0.45 | 0.55 | 645 | 7629 | 2936 | 724 | 166 |
| 12200 | 0.09 | 0.3 | 0.6 | 809 | 7459 | 2939 | 849 | 144 |
| 12200 | 0.07 | 0.35 | 0.6 | 646 | 7765 | 2805 | 839 | 145 |
| 12300 | 0.1 | 0.5 | 0.55 | 1209 | 7463 | 2885 | 594 | 149 |
| 12300 | 0.1 | 0.45 | 0.55 | 1054 | 7409 | 2951 | 726 | 160 |
| 12300 | 0.08 | 0.45 | 0.55 | 937 | 7498 | 2950 | 735 | 180 |
| 12300 | 0.06 | 0.35 | 0.6 | 584 | 7873 | 2889 | 793 | 161 |
| 12400 | 0.1 | 0.5 | 0.55 | 1138 | 7618 | 2897 | 603 | 144 |
| 12400 | 0.08 | 0.5 | 0.55 | 959 | 7720 | 2945 | 628 | 148 |
| 12400 | 0.08 | 0.35 | 0.6 | 715 | 7817 | 2881 | 832 | 155 |
| 12500 | 0.07 | 0.45 | 0.55 | 748 | 7850 | 2970 | 782 | 150 |
| 12800 | 0.1 | 0.35 | 0.6 | 973 | 7928 | 2928 | 827 | 144 |
| 12900 | 0.08 | 0.35 | 0.6 | 822 | 8152 | 2969 | 813 | 144 |

*N is the total number of genes in early vertebrates before 2R. P0 is the probability for a gene to be lost in the genome if there is only one homologous copy of this gene left in the genome, and 1-P0 is the threshold with which a random number generated during the simulation study is compared. P1 is the probability for a random gene to lost in the genome during the first round of whole genome duplication if there are more than one homologous copy. P2 is the probability for a random gene to lost in the genome during the second round of whole genome duplication if there are more than one homologous copy. N0 is the number of genes with no copy left in the genome. N1, N2, N3 and N4 are specifically the numbers of genes with 1, 2, 3, 4 copies left in the genome. Note this table just lists combinations of parameters that fit our reconstruction data.*

predicted Ohnologs are unsupported by our research, extra branches could be added to Singh et al's predicted Ohnolog families based on our research. Only 46.7% (818/1760) of Ohnolog families predicted by Singh et al's research have the same number of branches with synteny evidence as compared with the corresponding duplication nodes in our research. 41.4% (729/1760) Ohnolog families contain genes that are unsupported with synteny evidence in our research. 13.35% (235/1760) Ohnolog families miss genes/branches that have synteny evidence based on our research. Note the last 2 conditions may overlap and be repeatedly counted for some Ohnolog families. The Ohnologs inferred from our analysis are included in Supplemental Material Part3.

**Table 3 Comparison of Ohnologs groups from our research and Singh et al.'s research**

| Family sizes in Singh et al's rearch | # of families in consistent with PANTHER phylogenetics | # of families that have same phylogeny and synteny evidence compared with our research | # of families with genes unsupported by our research | # of families that are incomplete: missing genes/branches that have synteny evidence in our research |
|---|---|---|---|---|
| 2 | 1175 | 546 | 503 | 138 |
| 3 | 438 | 207 | 174 | 65 |
| 4 | 135 | 62 | 47 | 27 |
| >=4 | 12 | 3 | 5 | 5 |

**Potential misidentification of the time of duplications in PANTHER**

For this study, we extract genes from duplications only at early vertebrates, as inferred from the PANTHER phylogenetic trees. Thus, we are concerned about potential misidentification of the time of duplications: duplications that happen after early vertebrates such as fish-specific, tetrapod-specific, amniotes-specific duplications as well as duplication that happen before early vertebrates such as chordate-specific duplication. We therefore explored the extent of potential misidentification.

In PANTHER, the duplication time periods of genes are predicted through homologs in different species. For example, we predict a gene is duplicated at early vertebrates if multiple copies of this gene's paralogs are present in both fish and land vertebrates. If only one paralog is found in fish but multiple copies are found in land vertebrates, then we can only predict the duplication periods to be before the divergence of land vertebrates. However, it is possible that one or more copies of the paralogs are lost in ancestral fish, and thus the duplications have happened at early vertebrates instead of early land vertebrates. This may provide an explanation for the relative small number of duplications at early vertebrates. (Figure 3) To test this possibility, we looked at the genes that have been further duplicated at periods later than early vertebrates out of the whole sets of genes. For each duplication node at branches younger than early vertebrates (node A), if genes in 2 or more branches of this duplication node show synteny evidence with genes from other branches of the duplication node at early vertebrates (node B), it is extracted out together with the number of branches of node B and the number of branches that show evidence of synteny including the branch with node A.

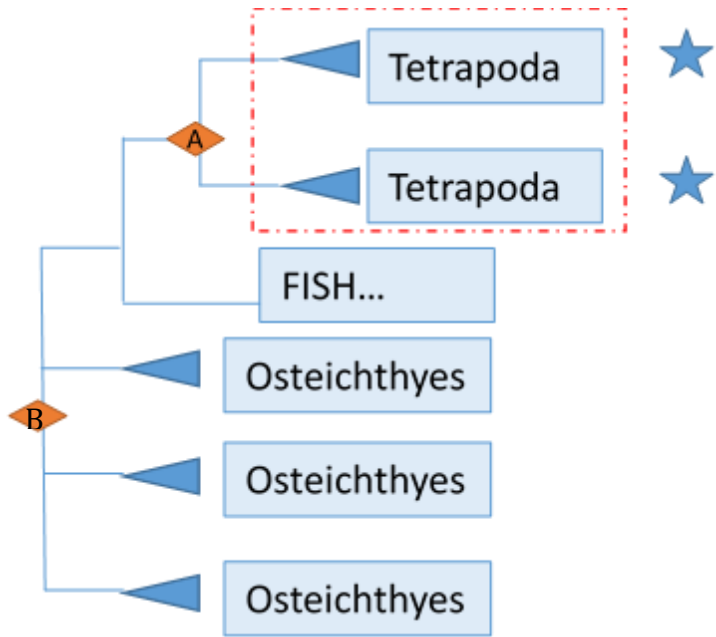

**Figure 3 Illustration of potential Misidentification of the time of duplications**

*This figure shows a simplified part of a phylogenetic gene tree. The blue box followed by a blue triangle represent a gene taxa at the specific period of the speciation name in the box. For example, Tetrapoda taxa includes genes from tetrapod species. The orange diamonds labeled with A and B are duplication nodes. Note they have 2 or more branches which are the gene taxa mentioned above. Node A is a duplication node at Tetrapoda, and Node B is a duplication node at Osteichthyes. We looked for gene trees with similar structures as shown in this figure: with duplication nodes at Osteichthyes, and one or more child branches have further duplication nodes linked to periods later than early vertebrates. For each duplication node at branches younger than Osteichthyes (node A), if genes in 2 or more branches (in red boxes) of this duplication node show synteny evidence with genes from other branches of the duplication node at Osteichthyes (node B), it is extracted out and used for later analysis.*

Table 4 lists the synteny evidences of genes that both have been duplicated at early vertebrates and are further duplicated at more recent time periods. From the results, we see many duplication nodes at early vertebrates with 4 or more branches. Thus, we could neglect the possibility of misidentification of duplication time periods due to gene losses, as it would yield many duplication nodes at early vertebrates with 4 or more branches. From another point of view, even if these tree structures were indeed misidentified, these tree structures only accounted for a very small percentage and would have little effects on our conclusions. Overall only 8.9% of the duplication nodes at early vertebrates have "node A" like duplication nodes at more recent time periods; and the descendant branches of these later duplication events have synteny evidence compared with branches of duplication nodes at early vertebrates.

**Table 4 synteny evidences of genes that have been duplicated at early vertebrates and are further duplicated at more recent time periods**

| Number of branches of a duplication node at Osteichthyes | Total number of these duplication nodes at Osteichthyes with a "node a" in any one branch | Total number of duplication nodes at Osteichthyes with a least one branch show evidence of synteny | Percentage of column 2 to column 3 |
|---|---|---|---|
| 2 | 71 | 899 | 0.078977 |
| 3 | 46 | 474 | 0.097046 |
| 4 | 23 | 192 | 0.119792 |

## Are duplication events that happen at later periods than early vertebrates due to whole genome duplications?

We extracted all the duplication nodes that have happened after early vertebrates and their descendant leaf genes to examine the evidence of synteny using the exact protocol for synteny evidence examination for the early vertebrate branch. Out of a total of 728 duplication nodes later than the 2R hypothesis with some evidences of synteny, 170 (23.3%) have multiple branches with evidence of synteny. Thus, our results do not support whole genome duplications at later periods than early vertebrates, but the duplication events at these younger periods are likely results of small-scale regional duplications.

## History and future of the 2R hypothesis

Ohno presented the first version of the 2R hypothesis based on relative genome sizes and isozyme analysis almost 50 years ago(Ohno 1970b). He suggested that ancestral fish or amphibians had undergone at least one and possibly more cases of "tetraploid evolution". He later added to this argument the evidence that most paralogous genes in vertebrates do not demonstrate genetic linkage. The 2R hypothesis saw a resurgence of interest in the 1990s. Gene mapping data of human and mouse revealed extensive paralogous regions. And the discovery of 4 HOX gene clusters in separate chromosomes in Human and mouse, in contrast to only 1 HOX gene cluster in amphioxus provided strong evidence the 2R hypothesis. The rapid increase of genomic data has thus provided more evidences of more paralogons within which gene duplications were dated to early vertebrates.

Rapid loss of genes right after the whole genome is widely accepted as an explanation for the lack of synteny evidences in extant species. The losses rate is estimated to be above 90% by some researchers. The hypotheses raised objections to the 2R arguing that the gene duplications did not happen in evolutionary history and thus would not lead to massive subsequent losses. Our gene gain and loss model at early vertebrates limits the loss rate to a reasonable range of 45-55%.

Objections to 2R hypothesis have come mainly from the phylogenic analysis of the genes in paralogons. It was argued by Hugh et al and Abbasi et al that paralogs A-D generated by two rounds of WGD should display the tree topology (AB) (CD) (Abbasi 2015; Friedman and Hughes 2001, 2003), while a substantial majority of genes in paralogons do not have this pattern. However, other researchers argued that incongruent tree topologies can be explained by two waves of genome doubling occurred in close succession, paralogs had different evolutionary rates, resolving power of phylogenetic trees is not sufficient, recombination could have occurred between the closely related chromosomes, genome duplication occurred through hybridization between species (Furlong and Holland 2002; Kasahara 2007; Lynch and Wagner 2009; Panopoulou and Poustka 2005).

With the results of this analysis, we could safely conclude that many segmental gene duplications have happened at early vertebrates, as these "segmental" duplications together span more than 50% of all genes, the whole genome duplications are the most likely explanation. However, several alternatives of 2R hypothesis are still possible. The first alternative hypothesis is that part of genome (multiple chromosomes) instead of the whole genome undergoes 2 rounds of genomes. A slightly different version of this hypothesis is that the first round of duplication is whole genome duplication but the second round of duplication happens only to part of the genome. The second alternative hypothesis is that there are a series of regional duplications at early vertebrates.

We could not eliminate these 2 hypotheses because (1) there is no evidence that all genes that were present in vertebrate ancestors had been duplicated in history. However, we could not exclude the possibility that these genes hadn't been duplicated at all, thus wouldn't have lost afterwards. (2) There is no evidence to link all paralogons together, thus duplications of these paralogons may not have happened at the exact same time periods. Duplication nodes of more than 4 branches exist, thus tandem duplications or small scale regional duplications may also have a role in duplications at early vertebrates.

While possibilities of the alternatives still exist, there have no convincing evidences to support these alternatives. 2 rounds of whole genome duplications are still the best explanation for the emergence of early vertebrates. In the future, with the availability of more and more vertebrate genomes, we would potentially be able to reconstruct evolutionary history of genes sequences change and genes order rearrangements with better accuracy.

**Conclusions**
In this study, we have summarized the most complete evidences of gene order based genome homology from 17 vertebrates and a close-related invertebrate Florida lancelet by extracting the genes that are duplicated at or after early vertebrates.

We have found extensive duplications that have happened at early vertebrates, and more than half of the identified duplication nodes at early vertebrates show some evidence of synteny. This is by far the most comprehensive evidence for 2R hypothesis based on strict gene order criteria. Based on our results, we can conclude that extensive long segmental gene duplications have happened at early vertebrates. Besides, the duplications that have happened at early vertebrates have a distinct pattern compared with duplications that have happened at other time periods in the sense of number of duplication, the number of branches of duplication nodes and how many of the branches show evidence of synteny. We have also identified 1740 Ohnologs groups that show evidence of synteny, and 1286 potential Ohnologs groups without synteny evidence. The lack of evidence could be possibly explained by gene losses and chromosomal rearrangements.

We also fit a gene loss model after whole genome duplications with reconstructions of ancestral genes assembled from PANTHER phylogenetic trees. We estimate that there are about 12000 genes in early vertebrates, the probability of a random gene gets lost after the first round of whole genome duplication is around 0.45, the probability of a random gene gets lost after the second round of whole genome duplication is around 0.55, and the probability of a random gene get lost if there is only one copy of this gene left in genome is 0.09. These parameters fit well with general expectations under the 2R model and probability for losses, with the second round having a higher probability for loss, and the loss rate of the copy left in genome being relatively small.

In summary, this research revisited the 2R hypothesis from a new perspective, which combines phylogenetic and gene order based genome homology. This research provides a strong evidence for 2R hypothesis, and may provide useful information for further insights in vertebral evolution.